\RequirePackage{fix-cm}
\documentclass[twocolumn]{svjour3} 
\smartqed 
\usepackage{graphicx} \journalname{Plasmonics}

\begin{document}

\title{Improving the sensitivity of fiber optical surface plasmon
  resonance sensor by filling liquid in a hollow core photonic crystal
  fiber}

\author{Zhixin Tan \and Xuejin Li \and Yuzhi Chen \and Ping Fan }


\institute{  Zhixin Tan \at Xuejin Li \at Yuzhi Chen \at 
             Shenzhen Key Laboratory of Sensor Technology,  \\
             Nanhai Ave 3688, Shenzhen, 518060, P. R. China \\
              Tel.: +86-0755-26538886           \\
              \email{tanzhxin@szu.edu.cn}      \\ 
              \email{lixuejin@szu.edu.cn}     \\ 
        \and 
              Zhixin Tan \at  Xuejin Li \at Yuzhi Chen \at 
              Shenzhen Engineering Laboratory for Optical Fiber Sensors and Networks, Shenzhen, 518060, P. R. China \\
        \and   Xuejin Li \at Yuzhi Chen \at
              College of Electronic Science and Technology, Shenzhen University, Shenzhen, 518060, P. R. China \\
        \and  Zhixin Tan \at Ping Fan \at 
             College of Physics Science and Technology, Shenzhen University, Shenzhen, 518060, P. R. China \\ 
             \email{fanping@szu.edu.cn} \\
}

\date{Received: date / Accepted: date}

\maketitle

\begin{abstract}
Inspired by the classic theory, we suggest that the performance of a
D-shape fiber optical surface plasmon resonance (SPR) sensor can be
improved by lowing down and tilting the fiber core mode. To
demonstrate this, we propose a novel fiber SPR sensor based on a
hollow core photonic crystal fiber with liquid mixture filled in
the core. The fiber sensor design involves a side-polished fiber with
gold film deposited on the polished plane and liquid filling. Our
numerical simulation suggests that by tuning the ratio of components
in the liquid mixture, the maximum predicted sensitivity of our model
for an aqueous analyte ($n$=1.33 ) will be over 6450 nm/RIU, which is
competitive with fiber chemical sensing. This design optimization method
may lead the way to an ultra-high sensitivity fiber optical biosensor.



\end{abstract}

\section{Introduction}
\label{intro}

Fiber optical SPR sensors have attracted much interest in recent years
for their compactness and \textit{in situ} sensing capability
~\cite{Yanase2010Developmentopticalfiber,Srivastava2012HighlySensitivePlasmonic,Slavik2002miniaturefiberoptic,Delport2012Realtimemonitoring,Verma2013FiberopticSPR,Kanso2008SensitivityOpticalFiber}. This
fiber sensor relies on a plasmonic wave, stimulated by the evanescent
field leaked from the fiber core, to sense biochemical
interactions. There are several ways to leak an evanescent field for
chemical sensing. One popular design involves integrating the sensing
surface on the inner walls of holes in the
fiber~\cite{Hassani2007Designcriteriamicrostructured,Hassani2006Designmicrostructuredoptical}. These
gold film coating holes are located near the fiber core, which
facilitates a direct excitation of plasmonic waves. This closed-form
all-in-one design reduces the consumption of sample volume and has
high
sensitivity~\cite{Popescu2012Powerabsorptionefficiency,Shuai2012multicoreholey}.
However, the fabrication of such a sensor is challenging, especially
the metal film deposition inside the holes. Moreover, with this sensor
the experimentalist lacks an auxiliary method to check the metal film
and the self-assembled bio-molecular film. Another approach to leak an
evanescent field is using post-processing, for example, by drawing or
side-polishing the fiber. A fiber with a diameter on the
order of the wavelength of visible light is ready to leak an
evanescent
field~\cite{Monzon-Hernandez2004OpticalFiberSurface,Tong2006Photonicnanowiresdirectly};
however, this drawing process will produce a highly fragile fiber
sensor which limit their application. A robust fiber sensor can be
fabricated by side-polishing an finished
fiber~\cite{Chen2011Sidepolishedfiber,Lo2011Ultrahighsensitivitypolarimetric,Tian2012AllsolidD}. This
method has the advantage of simplicity, as well as allowing good
control over metal film deposition. The planarity of the sensor region
allows the direct adoption of techniques from existing prism SPR
technology. In Ref.~\cite{Tian2012AllsolidD} the authors propose a
side polished fiber optic SPR sensor based on all-solid photonic crystal fiber. Since the
mechanical characteristics of the core and the cladding layer are
identical, the fabrication is simplified. The major problem with this
side-polishing sensor is that the sensitivity is limited in comparison
with the closed-form design. This problem should be addressed with
theoretical guidance.  After studying the SPR theory, we suggest that
decreasing and tilting the fiber core mode index is one way to improve
the sensor performance.

In this paper, we propose a novel SPR sensor based on a 
side-polished hollow core photonic crystal fiber. The design involves precise polishing of the base
fiber to form a uniform planar surface, upon which is deposited a gold
film; in addition, the hole along the center of the fiber is filled
with a
liquid~\cite{Xuejin2009AnalysisTemperatureSensing,Yu2010Somefeaturesphotonic},
the composition of which can be chosen to tune the core refractive
index. The propagating light will be confined in the fiber core,
mainly in the liquid-filled region. By changing the ratio of liquid
mixture, we tuning the refractive index of the fiber liquid core. The
numerical model will uncover how the filling liquid affects the
coupling of the core mode and the plasmonic mode and moves the phase
matching point, which is the fundamental to improve the sensor
performance, as characterized by its sensitivity.

This article is organized as follows. In section 2, we discuss the
model design and numerical methods. A specific application of the
model to the case of a fiber SPR sensor is illustrated in section
3. In section 4 we assess a serial fiber optical SPR sensor with
various choices for the central liquid filling, and discuss the
relationship between the core index and the sensitivity. Then a
summary in section 5 will end the paper.

\section{Model and sensor design}
\label{sec:model}
\begin{figure}
  \centering 
  \includegraphics[width=0.48\textwidth]{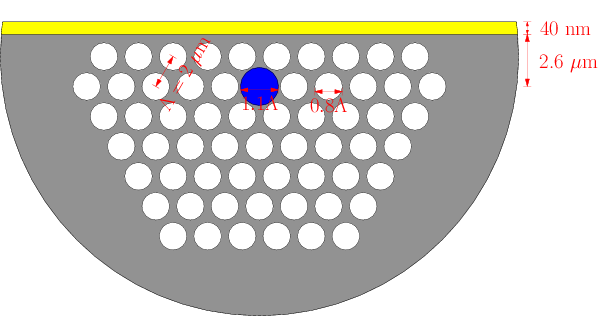}
  \caption{Schematic of the proposed fiber optical SPR sensor.}
  \label{fig:scheme}      
\end{figure}

A schematic of the fiber optical SPR sensor design is shown in
Fig.~\ref{fig:scheme}. The base fiber is a five-layer air-hole
photonic crystal fiber with a large hole in the center. The pitch of
the hexagonal lattice is $\Lambda = 2 \ \mu \mathrm{m}$. The diameter of
the air holes in the cladding layer is $0.8\Lambda$, and that of the
larger central hole is $1.1 \Lambda$. The dimension of the central 
core will influence the light confinement.To fabricate a sensor,
the fiber should be carefully polished so as not to impinge
on the underlying layer of holes and to ensure a smooth, planar
surface. This post-processing may be accomplished by laser
micro-fabrication. The height of the polished plane above the fiber
center is fixed at $2.6 \ \mu \mathrm{m}$, and the plane is coated with a gold
film 40 nm thick.

In the model, the refractive indices of silica and air are set as 1.45
and 1.0 respectively. The complex relative permittivity of gold is
treated using the Drude model [5]. The refractive index of the liquid
filling the central hole is variable. A convenient choice for the
filling liquid is a mix of purified water and glycerin; varying the
ratio of these two components allows the refractive index of the
mixture to be tuned within the range 1.33 to 1.47. The dispersion of 
the liquid mixture is ignored now and will be discussed later. The analyte is
located on the polished plane and has a refractive index of 1.33,
corresponding to an aqueous environment. Symbol $n_a$ represents the
refractive index of analyte, which is also the environmental refractive
index (RI) to be sensed.

A full vectorial FEM is used to solve the electromagnetic mode of the
fiber sensor~\cite{Pomplun2007Finiteelementsimulation}. The solution
in mode analysis is the electromagnetic field propagating in the
out-of-plane direction:
\begin{equation}
 E(r, t) = Re(\widetilde{E}(r) e^{i\omega t + \lambda z})
\end{equation}
The parameter $\lambda$ is a complex number defined as $\lambda =
-(\delta_z + i \beta)$ , where $\delta_z$ represents damping along the
z-axis and $\beta$ is the propagation constant. The effective mode
refractive index is defined as $n_{\mathrm{eff}}=i \lambda/k_0$ ,
where $k_0$ is the wave vector of the light in vacuum. The attenuation
per meter in dB is correlated to the imaginary part of
$n_{\mathrm{eff}}$ according to
\begin{equation}
   L_c = 20\log10 (e) \mathrm{Im}(k_0 n_{\mathrm{eff}})
\label{eq:loss}
\end{equation}
Here `e' is the base of the natural logarithm, $L_c$ means the loss of
the core mode which is measurable to quantify the performance of the
fiber sensor.

\section{Fiber optical SPR sensor with filling liquid of RI=1.39}
\label{sec:139}

\begin{figure*}
  \centering \includegraphics[width=0.9\textwidth]{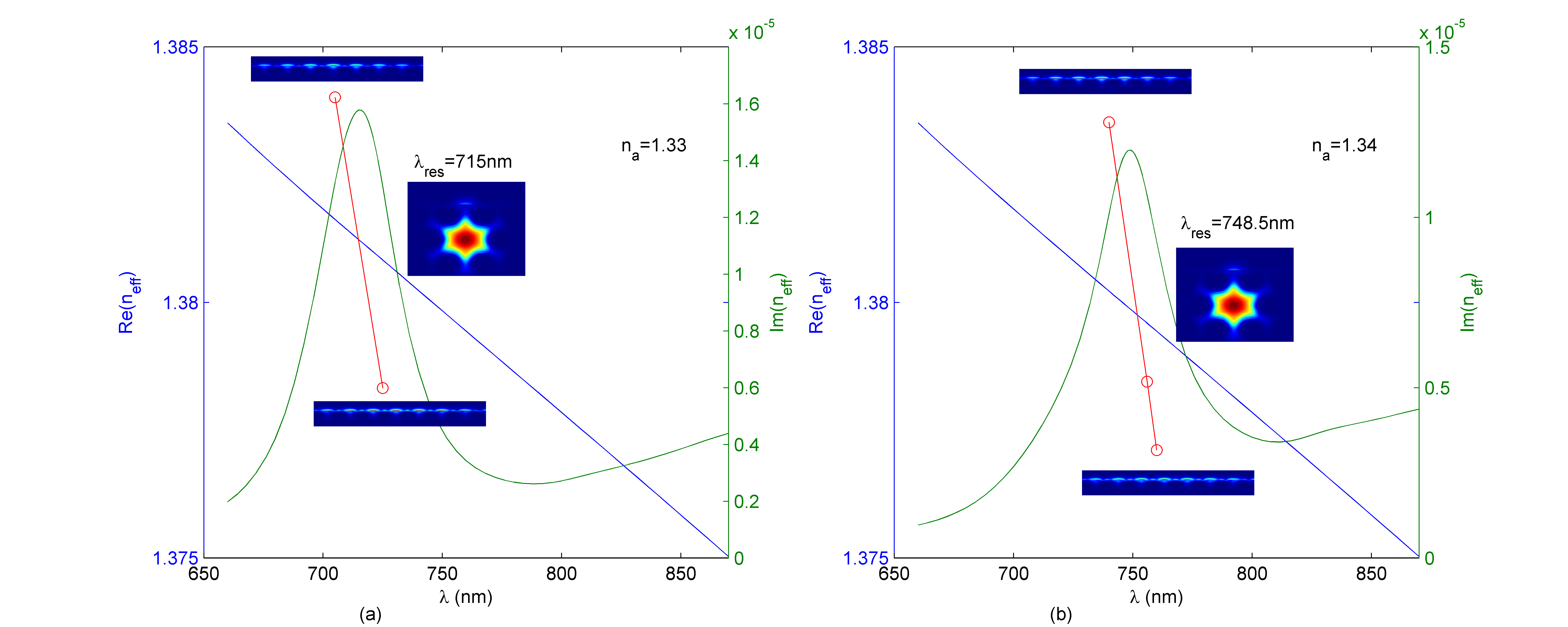}
  \caption{Numerical results for a fiber optical SPR sensor with
    filling liquid of RI = 1.39 and environmental refractive index
    $n_a$= 1.33 (left) and 1.34 (right). The blue line is the real
    part of the effective refractive index ( $n_{\mathrm{eff}}$ ) of
    the fiber sensor with respect of the input wavelength, and the
    green line is the imaginary part of $n_{\mathrm{eff}}$. The red
    line is the plasmonic wave. The resonance is characterized by the
    imaginary part of $n_{\mathrm{eff}}$ with a peak at the crossing
    point of the plasmonic mode (the red line) and the guide mode (the
    blue line).}
\label{fig:couple139}   
\end{figure*}

We simulated plasmonic wave stimulation on the sensor surface with the
fiber sensor operating in the lp01 mode. In this section we focus on
the case of a filling liquid refractive index of
1.39. Fig.~\ref{fig:couple139} depicts the stimulation of a plasmonic
wave on the metal surface with environmental refractive index $n_a$ =
1.33 (left) and 1.34 (right). In each sub-figure, the blue line is the
real part of the effective refractive index ($n_{\mathrm{eff}}$) of
the core guide mode as a function of the input wavelength, the green
line is the imaginary part of $n_{\mathrm{eff}}$ , and the red line is
the effective refractive index of the surface plasmon wave. The
resonance is characterized by the imaginary part of $n_{\mathrm{eff}}$
with a peak at the crossing point of the plasmonic mode (the red line)
and the guide mode (the blue line).  The upper and lower insets are
the electric field distribution of the plasmonic mode at the
corresponding red point.  The central insets show the spatial
distribution of the field in the fiber sensor near the resonance
wavelength. In the central insets, plasmonic waves are stimulated and
propagate along the surface above the fiber center.

\begin{figure}
  \includegraphics[width=0.5\textwidth]{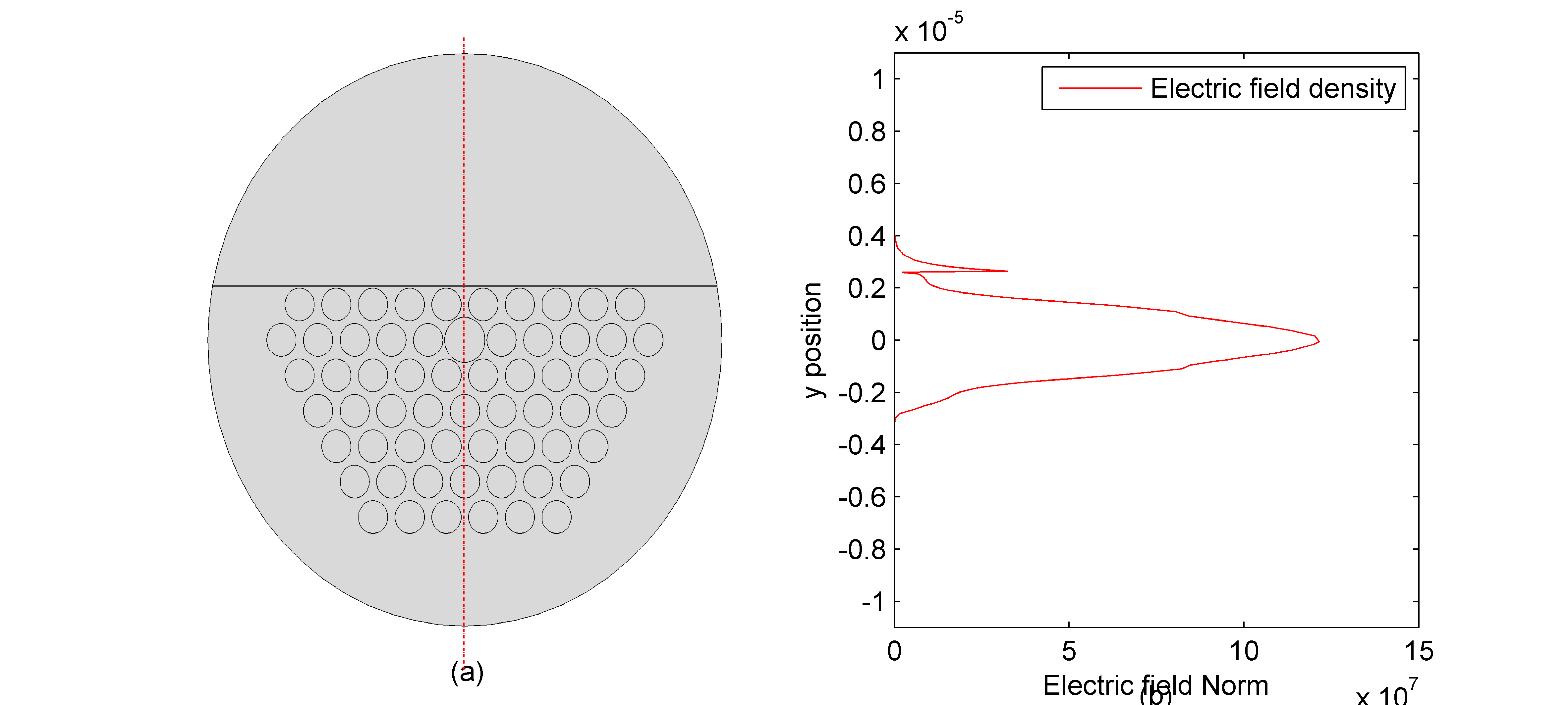}
  \caption{Electric field along the y-axis (red line in the schematic
    on the left). When y = $2.6~\mu \mathrm{m}$, at the height of the metal
    surface, the decay of the field is interrupted and a strong peak
    occurs corresponding to the plasmonic mode.}
  \label{fig:emline}   
\end{figure}

To study the resonance in detail, we investigate the electric field
distribution when the environmental refractive index equals
1.33. Since only the component of the electric field normal to the
metal surface can stimulate the plasmonic wave, in
Fig.~\ref{fig:emline} we plot the variation of the electric field
magnitude at the resonance wavelength with position along the vertical
(y) axis. Close to the center, the electric field density decreases
roughly symmetrically with radial distance. However, at y = 2.6 $\mu
\mathrm{m}$, corresponding to the plane of the metal surface, a sharp
peak in the electric field is apparent. This is the result of surface
plasmon resonance, for which the electric field is significantly
enhanced by the plasmonic wave at the metal surface. As expected, the
field decreases exponentially with distance above the surface.

\begin{figure}
\includegraphics[width=0.45\textwidth]{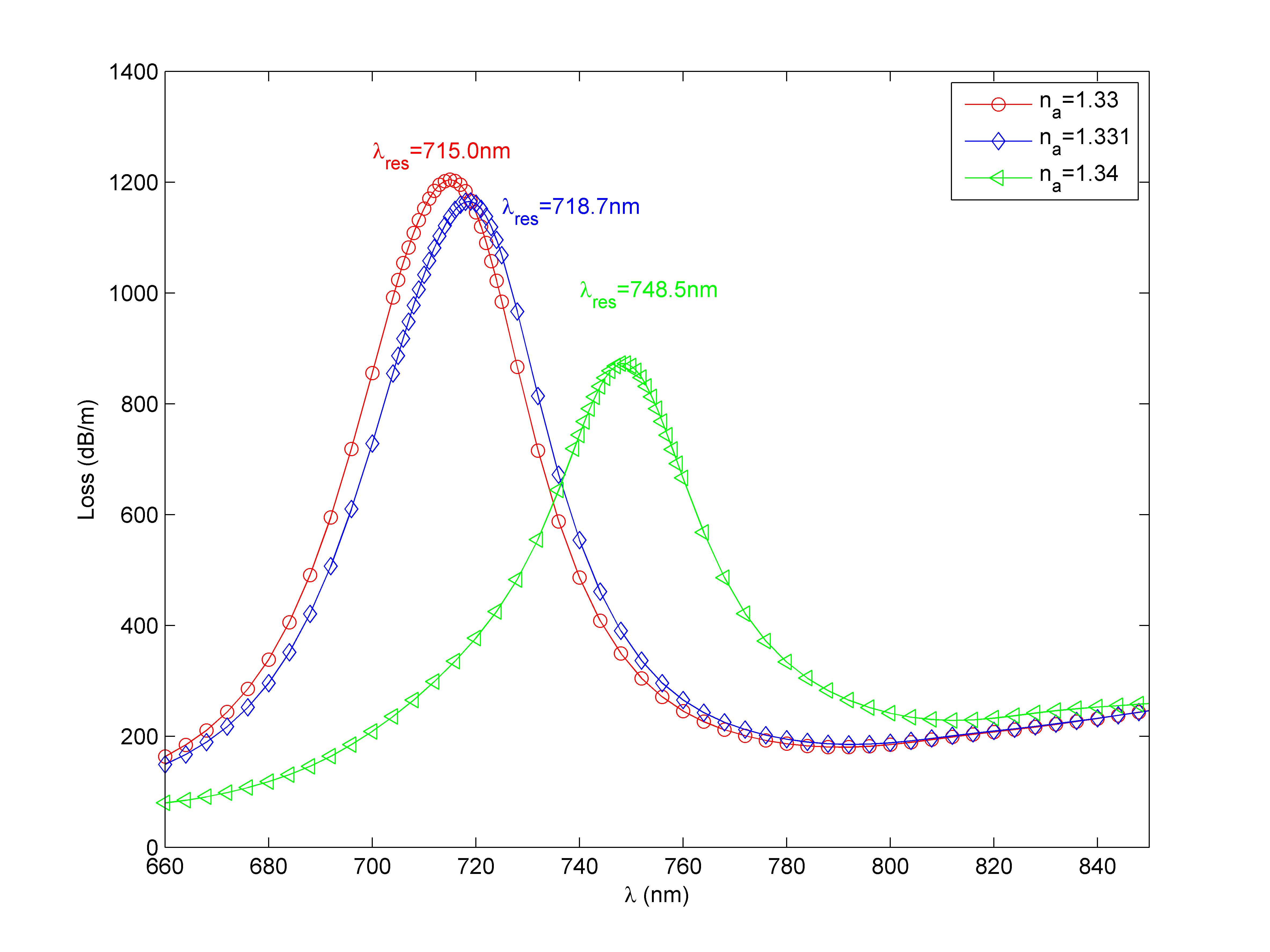}
\caption{Loss spectra and the resonance wavelengths of the fiber
  optical SPR sensor with environmental refractive index = 1.33, 1.331
  and 1.34, respectively.}
\label{fig:loss139}   
\end{figure}

The sensitivity of the fiber optical SPR sensor is evaluated by
wavelength interrogation. The sensor relies on shifts in the resonance
wavelength induced by changes in the refractive index at the fiber
surface, which may arise from mass variations on the surface
associated with biological interactions. The sensitivity of a fiber
optical SPR sensor is thus defined as the ratio of the shift of the
resonance wavelength to the change of the environmental refractive
index:
\begin{equation}
 S_\lambda(\mathrm{nm}/\mathrm{RIU}) = \frac{ \delta{\lambda_{res} } }{\delta{n_a}}
\label{eq:sensitivity}
\end{equation}
Where $\delta \lambda_{res}$ is the shift of the resonance peak
position in nm, and $\delta{n_a}$ is the change of the environmental
refractive index. To estimate the sensitivity near $n_a$= 1.33, the
loss spectra for $n_a$ = 1.33, 1.331 ($\delta{n_a}$=0.001) and 1.34 (
$\delta{n_a}$ = 0.01) are shown in Fig.~\ref{fig:loss139}.

When $\delta n_a$ = 0.01 (0.001), the estimated sensitivity is about
3350 (3700) nm/RIU. The precision of these values is limited by the
resonant wavelength measurement, the resolution of which is
interpolated to get $\pm 0.1$ nm. This numerical result is already
much higher than previously reported values for similar models
[8]. Moreover, we demonstrate below that sensitivity much higher than this
value can be obtained by optimizing the choice of filling liquid.

From Eq.~\ref{eq:loss}, the resonance wavelength in loss spectra is
the same value as the peak position of the loss curve under wavelength
interrogation. Therefore, we can skip the loss calculation and just
focus on the coupling chart where the resonance wavelength can be read
from the crossing point of the plasmonic mode and the core
mode. Later, we will evaluate the sensitivity in this way based on the
pair of environmental index values $n_a = 1.33/1.34$.

\section{Investigation of varying the refractive index of the fiber SPR sensor filling liquid }

\begin{figure*}
  \centering
  \includegraphics[width=0.95\textwidth]{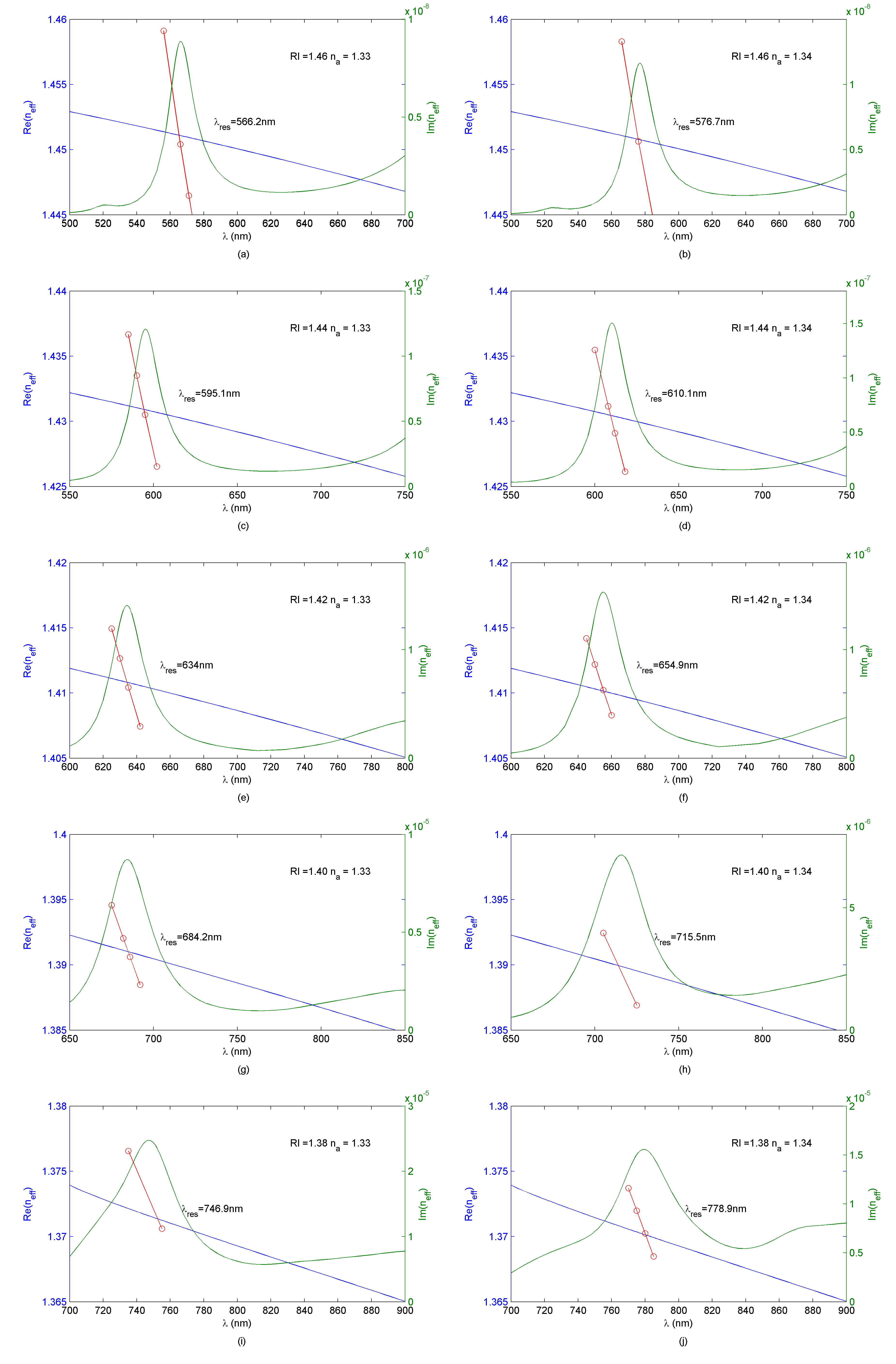}
\caption{Fiber SPR sensor filled with liquid mixture of (from top to
  bottom) RI = 1.46 1.44, 1.42, 1.40, and 1.38. The results on the
  left side are for environmental refractive index $n_a$= 1.33, and
  those on the right side for $n_a$= 1.34. Line colors are as in
  Fig.~\ref{fig:couple139}}
\label{fig:coupling5}   
\end{figure*}

To verify our hypothesis that the performance of a fiber SPR sensor is
directly affected by the core mode and to optimize the choice of
filling liquid, we present numerical modeling results in
Fig.~\ref{fig:coupling5} for different values of the filling liquid
refractive index: 1.46, 1.44, 1.42, 1.40, and 1.38.

The effective refractive index of the guide
mode is seen to vary with the refractive index of the filling liquid
correspondingly. This mode index denotes the index of the average
environment perceived by the propagating electromagnetic
field. Obviously, the mode index will be influenced by the choice of filling
liquid. Consequently, the phase matching point is moved. The resonance
occurs at the crossing points of the plasmonic modes and the core
modes, and the resonance wavelengths are labeled in
Fig.~\ref{fig:coupling5}.

\begin{figure}
  \includegraphics[width=0.5\textwidth]{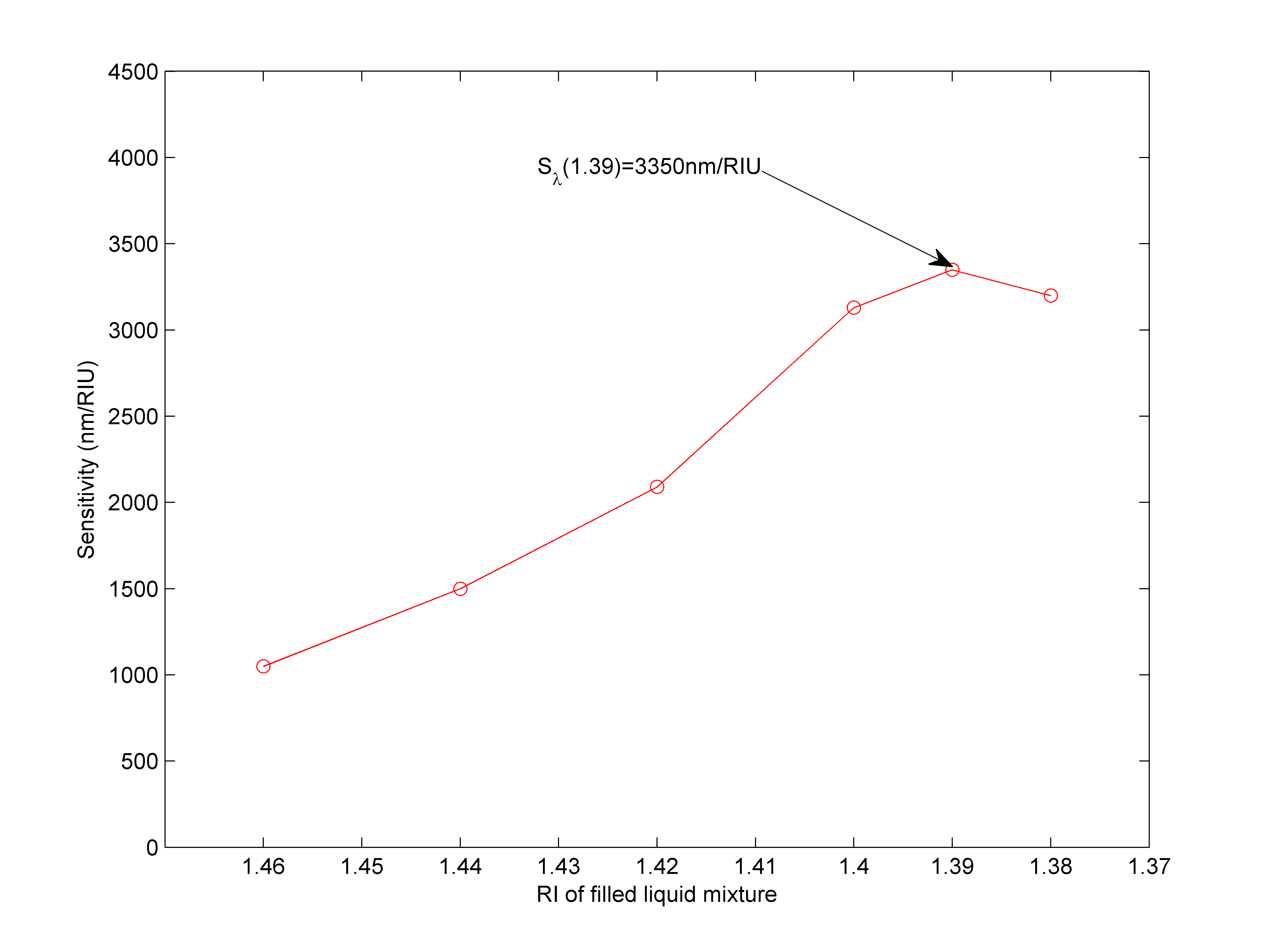}
  \caption{The sensitivities of the fiber optical SPR sensors with
    liquid of a decreasing RI.}
  \label{fig:sensitivity}   
\end{figure}

According to Eq.~\ref{eq:sensitivity}, the sensitivities of the fiber
optical SPR sensors with different filling liquids were evaluated by
comparing the resonance wavelengths for two different environmental
refractive indices (1.33 and 1.34). Fig.~\ref{fig:sensitivity} plots
the sensitivity as a function of the liquid refractive index. As the
liquid refractive index decreases (x-axis), the sensitivity of the
fiber SPR sensor increases, reaching a maximum when the filled liquid
RI = 1.39.

\begin{figure*}
  \centering \includegraphics[width=0.9\textwidth]{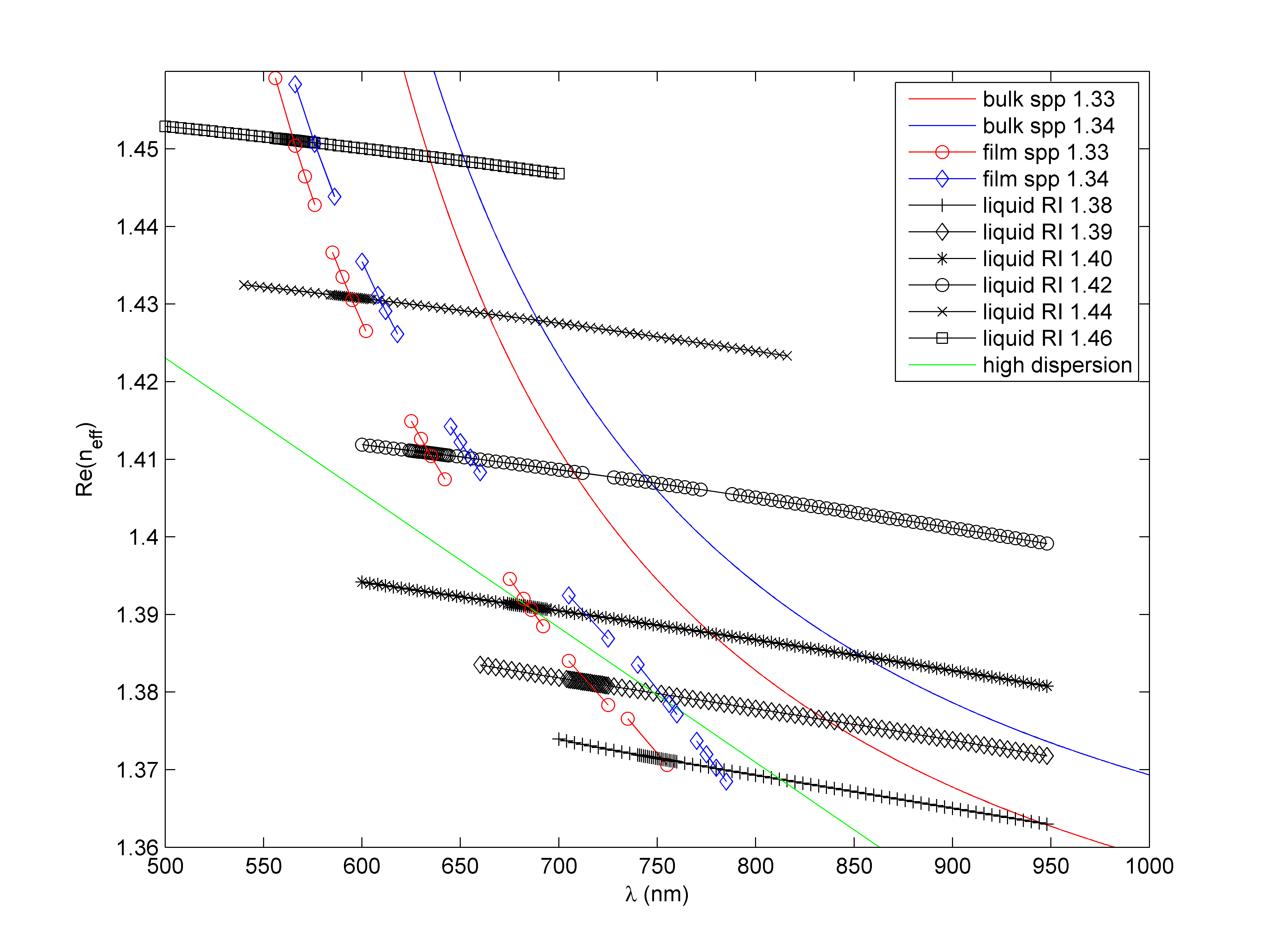}
  \caption{A comprehensive view of of the coupling of the plasmonic
    wave and the fiber core mode. Pair of red/blue solid lines,
    plasmonic modes from the semi-infinite plane model; pair of
    red/blue points/lines, plasmonic modes from numerical calculation
    of the fiber sensor models; black lines with markers, core modes
    from numerical calculations; green lines, effect of the dispersion
    on the core mode. }
\label{fig:allinone} 
\end{figure*}

To clarify the relation of the mode index and the sensitivity, in
Fig.~\ref{fig:allinone} we combine all the previous results in a sigle
simultaneous view. The black lines with symbols are the core mode for
various choices of the filling liquid refractive index.The solid blue
($n_a=1.33$) and red ($n_a=1.34$) lines correspond to the real part of
the plasmonic mode for a semi-infinite two-plane metal/dielectric
model
~\cite{Raether1988Surfaceplasmonssmooth,Pitarke2007Theorysurfaceplasmons}:,
in which the wave vector of the plasmonic mode $k_{\mathrm{spp}}$ is
given by
\begin{equation}
k_{spp} = \frac{k}{c} \sqrt{\frac{\epsilon_{Au} n_a^2}{\epsilon_{Au} +
    n_a^2} }
\label{eq:kspp}
\end{equation}
Where spp refers to ``surface plasmon polarition'', $\epsilon_{Au}$ is
the dielectric constant of gold, and $n_a$ is the dielectric constant
of environment, the dispersion of this medium is ignored now. The effective
 refractive index of the plasmonic mode is given as
\begin{equation}
n_{\mathrm{spp}} = \sqrt{\frac{\epsilon_{Au} n_a^2}{\epsilon_{Au} +
    n_a^2} }
\end{equation}
While the enviromental refractive index ($n_a$) decrease, the gap between these two
curves increases.  This indicates that if we maintain the plasmonic
mode while suppressing the core mode, we will lower the points at
which the plasmonic and core mode curves cross. These crossing points
correspond to the resonance wavelengths. The result is that the
difference in resonance wavelength between $n_a = 1.33$ and 1.34 will
increase. Accordingly, the sensitivity of the fiber SPR sensor will be
improved.

These solid curves are theoretical results for an idealized model,
with a different geometry to our sensor design. But the general trend
holds for the numerical solution of our fiber optical SPR sensor
model. The black lines with different markers in
Fig.~\ref{fig:allinone} are the core modes of the fiber optical
sensors with different choices of filling liquid, while the red and
blue points lines represent the numerical results for the plasmonic
modes with environmental refractive index =1.33 (red) and 1.34
(blue). Although these plasmonic mode points are results from
different sensor models with different core refractive indices, they
fall on a continuous curve. This is because the central hole is some
distance from the interface, so the influence of the liquid refractive
index on the plasmonic mode is weak. More importantly, the shape of
the curve mapped out by the numerically derived points is very similar
to the analytical curves, despite a horizontal shift that is due to
the different geometries of the analytical and numerical models: in
our fiber sensor design the plasmonic wave is stimulated on a gold
film, while in the analytical model the metal is a semi-infinite
plane. Since it is water and silica on each side of the gold film, the
sensor model is asymmetric, which result in no analytical
solution~\cite{Raether1988Surfaceplasmonssmooth,Pitarke2007Theorysurfaceplasmons}; and our FEM result
is the best choice for this asymmetric model.

As described above, the resonance is located at the crossing points of
the plasmonic modes and the core modes. The sensitivity of the fiber
SPR sensor is evaluated from the gap between these crossing points for
the red and blue plasmonic curves. In Fig.~\ref{fig:allinone}, the
gaps produced by the core mode cross with the plasmonic mode increase
gradually as the refractive index of the filling liquid (greater than
1.39) is decreased. These numerical results demonstrate that the
performance of the SPR sensor can be improved by lowering the fiber
core mode refractive index.

Our method for decreasing the effective refractive index differs from
those in earlier reports [4, 5]. In our design, the electric field is
confined to the liquid-filled region. That is why we can successfully
lower the crossing lines. However, while the RI of the filling liquid
decreases, the electric field confinement will be worse, because part
of the electromagnetic field will penetrate the high-refractive-index
silica wall; the FHWM of the resonance peak also grows (see
Fig.~\ref{fig:coupling5}). This may be why the curves of the plasmonic
modes are distorted and the sensitivity decreases for the case
RI=1.38.

Furthermore, we note that choosing a liquid with high dispersion will
significantly increase the sensor performance. We call this a
``titling'' method as it involves clockwise rotation of the fiber core
mode curve (Fig.~\ref{fig:allinone}). We estimate the sensitivity in
the dispersive case by observing where two neighbouring black lines
(representing two dispersive points) cross the red and blue plasmonic
curves, respectively. For example, if the refractive index of liquid
is 1.40 at wavelength 684 nm and drops to 1.39 at wavelength 748.5 nm,
we will obtain a sensitivity of 6450 nm/RIU, as illustrated by the
green line in Fig.~\ref{fig:allinone} (the green line). This is
remarkable because we can obtain a large increase in sensitivity
without change in the structure of the fiber sensor. Of cause, these
two points serves as an example of high dispersive liquid.

To address the question of how much dispersion is required to obtain
this effect, we use a simple model to estimate the required Abbe
number of the liquid. The dispersion curve is extrapolated linearly
based on the n=1.39 and 1.40 data points. The Abbe number is estimated
as:
\begin{equation}
 V_d = \frac{n_d-1 }{n_F - n_c} = 15.7
\end{equation}
where $n_d$, $n_F$ and $n_C$ are the refractive indices of the
material at the wavelengths of the d-, F- and C- spectral lines (587.6
nm, 486.1 nm and 656.3 nm) respectively. This requirement is
attainable, since methanol has an Abbe number of $V_d$ = 13.66 with
refractive index 1.3172 at 750nm, while ethylene glycol has a Abbe
number of Vd = 12.86 with refractive index 1.41856 at
750nm~\cite{refractiveindexinfo}. A mixture of methanol and ethylene
glycol in a certain ratio should be able to simultaneously satisfy the
optimal index and dispersion properties for achieving high
sensitivity. Because the dispersion of a liquid mixture is
complicated, we stop the expansion and emphasize the method.

\section{Conclusion}

By constructing a comprehensive visualization of mode coupling curves,
we elucidate the fundamental physics underlying the optimization of a
D-shaped fiber optical SPR sensor.  A fiber sensor design is proposed
based on an hollow core photonic crystal fiber with a liquid filled in the central
hole. While the filling liquid satisfies the optimal index and high
dispersion property simultaneously, the performance of the fiber
optical SPR sensor can be improved significantly. Apart from having
high sensitivity, this sensor design has the advantage of simplicity:
the planar geometry of the gold film means that film deposition and
verification should be straightforward, which is important for the
feasibility of sensor applications.

\begin{acknowledgements}
This work is supported by the National Science Foundations of China
under Grants (No.61275125), the Research Foundation for the Doctoral
Program of Higher Education of Ministry of Education (20124408110003),
the Province-Ministry Industry-University-Institute Cooperation
Project of Guangdong Province under Grant (No. 2010B090400328), the
Shenzhen Science and Technology Project, the Shenzhen Nanshan District
Science and Technology Project, the High-level Talents Project of
Guangdong Province.

\end{acknowledgements}

\bibliographystyle{spphys}
\bibliography{myfill}


\end{document}